\documentclass[english]{article}
\usepackage[LGR,T1]{fontenc}
\usepackage[latin9]{inputenc}
\usepackage[a4paper]{geometry}
\usepackage{color}
\usepackage{babel}
\usepackage{float}
\usepackage{units}
\usepackage{amsmath}
\usepackage{amssymb}
\usepackage{setspace}
\usepackage[unicode=true,pdfusetitle,
 bookmarks=true,bookmarksnumbered=false,bookmarksopen=false,
 breaklinks=false,pdfborder={0 0 1},backref=false,colorlinks=false]
 {hyperref}
\usepackage{breakurl}

\makeatletter

\DeclareRobustCommand{\greektext}{%
  \fontencoding{LGR}\selectfont\def\encodingdefault{LGR}}
\DeclareRobustCommand{\textgreek}[1]{\leavevmode{\greektext #1}}
\ProvideTextCommand{\~}{LGR}[1]{\char126#1}

\providecommand{\tabularnewline}{\\}

\newcommand{\lyxaddress}[1]{
	\par {\raggedright #1
	\vspace{1.4em}
	\noindent\par}
}

\@ifundefined{date}{}{\date{}}
\makeatother

\begin{document}

\title{\textbf{On the existence of quaternion-handedness for the spinor
solutions of the Dirac and Majorana neutrinos}}

\author{\textbf{B. C. Chanyal}\thanks{\textbf{Email:}\textcolor{blue}{{} bcchanyal@gmail.com }}
~\textbf{and Sayyam Murari }}
\maketitle

\lyxaddress{\begin{center}
\emph{Department of Physics }\\
\emph{G. B. Pant University of Agriculture and Technology, Pantnagar-
263 145, Uttarakhand, India}
\par\end{center}}
\begin{abstract}
This study explores the application of quaternionic left and right-handed
algebraic frameworks to discuss the neutrino-handedness behavior.
The generalized left and right-handed Dirac equations are expressed
in terms of quaternionic four-dimensional non-commutative algebra,
which yields helicity eigenspinors with definite spin states for Dirac
neutrino eigenstates. We establish the spinor states of energy and
momentum solutions for Dirac neutrinos and antineutrinos revealed
by the non-commutative nature of quaternionic algebra. Furthermore,
we expand the quaternionic Dirac spinor solutions to Majorana-like
spinor solutions, which represent the mass eigenstates for Majorana
neutrinos. By examining two-flavor neutrino oscillation, we express the
quaternionic energy and momentum eigenstates of Dirac and Majorana
neutrinos, as well as their oscillation probabilities. The estimated
quaternionic-based probability supports neutrinos' massive nature
and the presence of their right chiral components. The current quaternionic
theory has the benefit of identifying the distinct properties of left
and right-handed neutrinos, as well as anti-neutrinos, within a unified
framework.
\end{abstract}

\section{Introduction}

Neutrinos are elementary particles in the Standard Model (SM) of particle
physics that play a significant role in various physical phenomena
and the evolution of the universe. They contribute to weak nuclear
interactions and reveal physics beyond the Standard Model (PBSM) through
oscillations. Chirality is an important concept for neutrinos that
affects particle behavior in SM and PBSM. The Standard Model of particle
physics is a comprehensive model that includes all fundamental particles
and their interactions. The relativistic Dirac equation describes
the behavior of all fundamental fermions of SM, including neutrinos,
because their experimental evidence was verified at high energy scales.
When the wave function for the neutrino in the Dirac equation is broken
down into two left and right chiral components, we get two chiral
wave equations associated with its rest mass. Since only left-handed
neutrinos participate in the weak interaction, in view of SM. The
left chiral component of the wavefunction means that the neutrino's
rest mass is zero. However, neutrino oscillations have revealed that
they are substantial, albeit with extremely small masses in compared
to other fundamental fermions, leading to the possibility of right-handed
neutrinos. Nonetheless, they do not participate in the weak interaction
and are hence referred to as sterile neutrinos\textbf{ }\cite{key-1}.
Recent advances in high-energy physics necessitate the extension of
the SM \cite{key-2,key-3}, which continues to fuel ongoing research
and experiments. Furthermore, the type of neutrino mass remains unidentified
as the neutrino could be Dirac or Majorana neutrino. The Dirac equation
describes the Dirac neutrinos. Majorana \cite{key-4} suggested the
massive like neutrino, which has its own antiparticle, and Case \cite{key-5}
discovered their two component solutions, which also describes them
as non-parity conserving. However, the two Majorana spinors contradict
the lepton number and are Lorentz invariant \cite{key-6}. According
to Pontecorvo \cite{key-7}, neutrino wave functions involved in weak
interactions are a superposition of Hamiltonian eigenstates. Furthermore,
the oscillations of neutrinos with both Dirac and Majorana mass terms
were explored in \cite{key-8}. Using spinor solutions of the Dirac
equation to represent mass eigenstates, the neutrino oscillations
were explored \cite{key-9} using the wave packet formalism. The flavor
and spin oscillations of neutrinos in a constant magnetic field have
been addressed as well \cite{key-10}. The Majorana- like neutrino
oscillations have been clarified by adding another Majorana phase
to the mixing matrix \cite{key-11,key-12}. These studies give sufficient
background to incorporate both the Dirac equation and Majorana neutrinos
into the neutrino oscillation phenomenon. This represents why most
scientists consider neutrinos to be massive, regardless of how their
masses are.

Besides that, the role of higher-dimensional complexified division
algebra is crucial to explaining the characteristics of neutrinos'
left and right chiral components. In this direction, we propose a
quaternionic-based massive neutrinos theory using their transition
probabilities. The quaternionic algebra is a four dimensional extension
of complex numbers with non commutative properties \cite{key-13},
which shows an additional degree of freedom. This gives rise to quaternions
with left- and right-handedness due to the non-commutative property
\cite{key-14}. Additionally, there have been numerous successful
attempts to combine quaternionic algebra into relativistic quantum
physics. Conway \cite{key-15} developed the quaternionic rotational
operator to explain the Lorentz transformations, which are simply
rotations in four dimensions of spacetime. Similar attempts have been
made in \cite{key-16,key-17,key-18,key-19}, which show that the quaternionic
basis elements in both handedness may be adapted to Pauli spin matrices.
Moreover, the role of quaternionic algebra has been used to study
the dual magneto-hydrodynamics \cite{key-20}, quantum field equations
\cite{key-21}, rotational Dirac equations in electromagnetic field
\cite{key-22}, non-linear field equations \cite{key-23}, generalized
Klein paradox \cite{key-24}, behaviour of dyonic matter \cite{key-25},
non-relativistic quantum mechanics \cite{key-26,key-27}, generalized
Virial theorem \cite{key-28}, Gravito-Electromagnetism \cite{key-29},
particle in a relativistic box \cite{key-30}, electromagnetism \cite{key-31},
multifluid plasma equations \cite{key-32}, and the hydrodynamic two-fluid
plasma \cite{key-33}. Keeping in view the recent updates on quaternions,
in the present paper, we formulate the generalized Dirac equation
in terms of both left and right quaternions by connecting the gamma
matrices to the quaternionic basis elements. We have incorporated
the quaternionic valued mass for the dynamics of neutrinos. Quaternionic-based
energy and momentum spinor solutions are also addressed for neutrinos
flavors. We establish the quaternionic form of Majorana spinors in
both quaternionic handedness by equating Dirac's charge conjugate
spinors to themselves. Further, we propose the relationship between
the left and right quaternionic spinors and their role as mass eigenstates
to discuss the neutrino flavor quaternionic eigenstates, which are
eventually used to calculate the transition probabilities to analyze
the massive nature of neutrinos. This can find significance in the
in the role of right-hand neutrinos in various hidden theories, e.g.,
dark matter, dark energy, hierarchy problems, matter-antimatter asymmetry,
etc.

\section{Quaternions }

In the norm division algebras, complex algebra plays an important
role for various physical theories with satisfying commutative properties.
A complex number ($\mathbb{C}$) is defined as the set of all real
linear combinations of the unit elements $(1,\,i)$, such that $\mathbb{C}\,:\longmapsto\,\left\{ z=x+iy\,\mid\,(x,\,y\in\mathbb{R})\right\} \,,$
where $x$ indicates the real part while $y$ indicates the imaginary
part of a complex number $z$. The Euclidean scalar product of two
complex numbers denoted by $\mathbb{C}\times\mathbb{C}\longmapsto\mathbb{R}$,
is then expressed as
\begin{align}
\left\langle u,\,v\right\rangle \,= & \,\,Re\,(u\cdot\bar{v})\,\,=\,\,u_{1}v_{1}+u_{2}v_{2}\,,\label{eq:1}
\end{align}
where $\left(u,v\right)\in\mathbb{C}$. Also, the commutative property
satisfy the relation $\left\langle u,\,v\right\rangle =\left\langle v,\,u\right\rangle $,
which implies that the left-handed product is the same as the right-handed
product of any two complex variables. To distinguish the left and
right-handed products, we must apply the quaternionic algebra. A quaternionic
variable (or quaternion `$\text{\ensuremath{\mathbb{Q}}}$') is expressed
with the linear combinations of the unit elements $(e_{0},e_{1},e_{2},e_{3})$,
which is an extension of complex numbers, such that \cite{key-13},
\begin{equation}
\mathbb{Q}\,\,=\,\,z_{1}+z_{2}e_{2}\,\equiv\,\left(e_{0}q_{0}+e_{1}q_{1}\right)+\left(e_{0}q_{2}+e_{1}q_{3}\right)e_{2}\,=\,\,e_{0}q_{0}+\sum_{j=1}^{3}e_{j}q_{j}\,,\label{eq:2}
\end{equation}
where \textcolor{black}{$\left(q_{0},q_{1},q_{2},q_{3}\right)\in\mathbb{R}^{4}$.}\textbf{\textcolor{black}{{}
}}The quaternionic conjugate of equation (\ref{eq:2}) is expressed
by
\begin{align}
\bar{\mathbb{Q}}\,\,= & \,\,e_{0}q_{0}-\sum_{j=1}^{3}e_{j}q_{j}\,.\label{eq:3}
\end{align}
Now, using equations (\ref{eq:2}) and (\ref{eq:3}), we can further
define a quaternion in terms of scalar $S(q)$ and vector $\boldsymbol{V}(q)$
parts as 
\begin{align}
\mathbb{Q}\,\,=\,\,S(q)+\boldsymbol{V}(q),\,\,\forall\,\, & \left[S(q)\,=\,\frac{1}{2}(\,\mathbb{Q}\,+\,\bar{\mathbb{Q}}\,)\,,\,\,\boldsymbol{V}(q)=\,\frac{1}{2}(\,\mathbb{Q}\,-\,\bar{\mathbb{Q}}\,)\right]\,.\label{eq:4}
\end{align}
Since, the addition of any two quaternions is associative and commutative,
however in multiplicative property the quaternions is associative
but non-commutative. The multiplication of two quaternions $\left(\mathbb{Q}\circ\mathbb{R}\right)$
can be written as
\begin{align}
\text{\ensuremath{\mathbb{Q}}}\circ\mathbb{R}\,\,= & \,\,e_{0}\left(q_{0}r_{0}-q_{1}r_{1}-q_{2}r_{2}-q_{3}r_{3}\right)+e_{1}\left(q_{0}r_{1}+q_{1}r_{0}+q_{2}r_{3}-q_{3}r_{2}\right)\nonumber \\
+ & e_{2}\left(q_{0}r_{2}-q_{1}r_{3}+q_{2}r_{0}+q_{3}r_{1}\right)+e_{3}\left(q_{0}r_{3}+q_{1}r_{2}-q_{2}r_{1}+q_{3}r_{0}\right)\,,\label{eq:5}
\end{align}
where the quaternionic unit elements $(e_{0}\,,e_{1},\,e_{2},\,e_{3})$
are satisfied the following relations,
\begin{align}
e_{0}^{2} & \,\,=\,1\,,\,\,e_{j}^{2}=\,-1\,,\,\,e_{0}e_{j}=\,e_{j}e_{0}=e_{j}\,,\nonumber \\
e_{i}e_{j} & \,\,=-\delta_{ij}e_{0}+f_{ijk}e_{k}\,,\,\,\,(\forall\,i,j,k=1,2,3)\,.\label{eq:6}
\end{align}
Here $\delta_{ij}$ and $f_{ijk}$ are well known delta symbol and
the Levi Civita three-index symbol, respectively. Furthermore, the
commutation and the anti-commutation relations for the quaternionic
basis elements are expressed as
\begin{align}
\left[e_{i},\,\,e_{j}\right] & \,=\,2\,f_{ijk}\,e_{k}\,,\,\,\,\,\,\,\,\,\,\left\{ e_{i},\,\,e_{j}\right\} \,=\,-2\,\delta_{ij}e_{0}\,,\label{eq:7}
\end{align}
where the brackets have their usual meaning. Moreover, the quaternionic
valued of modulus $\mid\text{\ensuremath{\mathbb{Q}}}\mid$ and inverse
$\text{\ensuremath{\mathbb{Q}}}^{-1}$ can be defined as
\begin{align}
\mid\text{\ensuremath{\mathbb{Q}}}\mid\,= & \,\sqrt{q_{0}^{2}+q_{1}^{2}+q_{2}^{2}+q_{3}^{2}}\,;\,\,\,\,\,\,\text{\ensuremath{\mathbb{Q}}}^{-1}\,=\frac{\bar{\mathbb{Q}}}{\mid\text{\ensuremath{\mathbb{Q}}}\mid}\,.\label{eq:8}
\end{align}
While the norm of the quaternion can be expressed by the composition
law, such as, $\text{Norm}\left(\text{\ensuremath{\mathbb{Q}}}_{1}\circ\text{\ensuremath{\mathbb{Q}}}_{2}\right)\,=\,\left[\text{Norm}\,\,\text{\ensuremath{\mathbb{Q}}}_{1}\,\circ\,\text{Norm}\,\,\text{\ensuremath{\mathbb{Q}}}_{2}\right].$
The quaternion algebra represents a non-commutative division ring.

\subsection{Left and right-handed representation of quaternions}

\textcolor{black}{Generally, the left- and right-handed representations
are associated with particle spin and helicity. Also in quaternionic
formalism, we can write the left and right-handed quaternions \cite{key-14}
by using the multiplication rules of basis elements given by equation
(\ref{eq:6}). If the cyclic permutations of multiplication of basis
elements are positive, the nature of such type of product becomes
the left-handed quaternionic or ordinary quaternions. Thus, the multiplication
of quaternionic basis elements can be written as
\begin{align}
e_{j}^{2} & \,=\,-e_{0}^{2}=-1\,,\,,\,\,(\forall\,j=1,2,3)\nonumber \\
e_{i}e_{j} & \,=-\delta_{ij}+f_{ijk}e_{k}\,,\,\,\,\,\,\,\,\,\,\text{(Left handed representation)}\label{eq:9}
\end{align}
}such as $e_{1}e_{2}=e_{3}=-e_{2}e_{1}$, $e_{3}e_{1}=e_{2}=-e_{1}e_{3}$
and $e_{2}e_{3}=e_{1}=-e_{3}e_{2}$; while if the anti-cyclic permutations
of the multiplications of vector basis elements are positive, then
it becomes right-handed quaternions, such as\textcolor{black}{
\begin{align}
e_{j}^{2} & \,=\,-e_{0}^{2}=-1\,,\,,\,\,(\forall\,j=1,2,3)\nonumber \\
e_{i}e_{j} & \,=-\delta_{ij}-f_{ijk}e_{k}\,,\,\,\,\,\,\,\,\,\,\text{(Right handed representation)}\label{eq:10}
\end{align}
}where we have $e_{1}e_{2}=-e_{3}=-e_{2}e_{1}$, $e_{3}e_{1}=-e_{2}=-e_{1}e_{3}$
and $e_{2}e_{3}=-e_{1}=-e_{3}e_{2}$. In details, the multiplication
rules for left and right-handed quaternions can be summarized in Table
1(a) \& 1(b), respectively.
\begin{table}[H]
\begin{doublespace}
\begin{centering}
\begin{tabular}{c|cccc}
$\circ$ & \textbf{$e_{0}$} & \textbf{$e_{1}$} & \textbf{$e_{2}$} & \textbf{$e_{3}$}\tabularnewline
\hline 
$e_{0}$ & 1 & $e_{1}$ & $e_{2}$ & $e_{3}$\tabularnewline
$e_{1}$ & $e_{1}$ & $-1$ & $e_{3}$ & $-e_{2}$\tabularnewline
$e_{2}$ & $e_{2}$ & $-e_{3}$ & $-1$ & $e_{1}$\tabularnewline
$e_{3}$ & $e_{3}$ & $e_{2}$ & $-e_{1}$ & $-1$\tabularnewline
\hline 
\end{tabular}~~~~~~~~~~~%
\begin{tabular}{c|cccc}
$\circ$ & \textbf{$e_{0}$} & \textbf{$e_{1}$} & \textbf{$e_{2}$} & \textbf{$e_{3}$}\tabularnewline
\hline 
$e_{0}$ & 1 & $e_{1}$ & $e_{2}$ & $e_{3}$\tabularnewline
$e_{1}$ & $e_{1}$ & $-1$ & $-e_{3}$ & $e_{2}$\tabularnewline
$e_{2}$ & $e_{2}$ & $e_{3}$ & $-1$ & $-e_{1}$\tabularnewline
$e_{3}$ & $e_{3}$ & $-e_{2}$ & $e_{1}$ & $-1$\tabularnewline
\hline 
\end{tabular}
\par\end{centering}
\end{doublespace}
\begin{centering}
\begin{tabular}{cc}
\textbf{(a)} Left-handed multiplication & ~~~~~~~~\textbf{(b)} Right-handed multiplication\tabularnewline
\end{tabular}
\par\end{centering}
\centering{}\caption{Multiplication table for quaternionic basis elements}
\end{table}
It is important to point out that quaternion left and right-handed
multiplication shows no handedness behavior on multiplication by scalar
unit element $e_{0}$, whereas handedness behavior appears only on
multiplication by non-scalar units or non-equal basis elements. So,
for the left-handed quaternions, the multiplication of two quaternions
given in equation (\ref{eq:5}), can be simplified as

\begin{alignat}{1}
\left(\text{\ensuremath{\mathbb{Q}}}\circ\mathbb{R}\right)_{L} & \,\,=\,\,e_{0}\left(q_{0}r_{0}-\overrightarrow{q}\cdot\overrightarrow{r}\right)+\sum_{j=1}^{3}e_{j}\left[q_{0}r_{j}+r_{0}q_{j}+(\overrightarrow{q}\times\overrightarrow{r})_{j}\right]\,,\label{eq:11}
\end{alignat}
meanwhile, the right-handed multiplication becomes:
\begin{align}
\left(\text{\ensuremath{\mathbb{Q}}}\circ\mathbb{R}\right)_{R} & \,\,=\,\,e_{0}\left(q_{0}r_{0}-\overrightarrow{q}\cdot\overrightarrow{r}\right)+\sum_{j=1}^{3}e_{j}\left[q_{0}r_{j}+r_{0}q_{j}-(\overrightarrow{q}\times\overrightarrow{r})_{j}\right]\,,\label{eq:12}
\end{align}
where the subscripts\emph{ }`\emph{$L$}' and `$R$' denote left and
right-handed quaternions. In equations (\ref{eq:11}) and (\ref{eq:12}),
we observe that the order of multiplication affects the sign of the
cross product term that is included in pure quaternions, while the
other terms remain unchanged. This makes both left and right quaternionic
products non-commutative. Another important observation is that changing
quaternionic handedness is equivalent to changing the order of multiplication
for the quaternions, \emph{i.e.
\begin{align}
\left(\text{\ensuremath{\mathbb{Q}}}\circ\mathbb{R}\right)_{L} & \,\,=\,\,\left(\mathbb{R}\circ\text{\ensuremath{\mathbb{Q}}}\right)_{R}\,,\label{eq:13}\\
\left(\text{\ensuremath{\mathbb{Q}}}\circ\mathbb{R}\right)_{R} & \,\,=\,\,\left(\mathbb{R}\circ\text{\ensuremath{\mathbb{Q}}}\right)_{L}\,,\label{eq:14}
\end{align}
}where,
\begin{align}
\left(\mathbb{R}\circ\text{\ensuremath{\mathbb{Q}}}\right)_{L} & \,\,=\,\,e_{0}\left(r_{0}q_{0}-\overrightarrow{r}\cdot\overrightarrow{q}\right)+\sum_{j=1}^{3}e_{j}\left[r_{0}q_{j}+q_{0}r_{j}+(\overrightarrow{r}\times\overrightarrow{q})_{j}\right]\,,\label{eq:15}\\
\left(\mathbb{R}\circ\text{\ensuremath{\mathbb{Q}}}\right)_{R} & \,\,=\,\,e_{0}\left(r_{0}q_{0}-\overrightarrow{r}\cdot\overrightarrow{q}\right)+\sum_{j=1}^{3}e_{j}\left[r_{0}q_{j}+q_{0}r_{j}-(\overrightarrow{r}\times\overrightarrow{q})_{j}\right]\,.\label{eq:16}
\end{align}
Thus, the commutative property employs only when two quaternionic
variables change handedness with their reverse operation. Therefore,
the role of handedness is crucial in quaternionic formulation.

\section{Formulation of Left and right-handed Dirac equation for neutrino}

Neutrinos, the fundamental building blocks of matter, are massless
particles, and according to the standard model, only left-handed neutrinos
and right-handed antineutrinos exist. However, beyond the SM, recent
investigations \cite{key-34,key-35,key-36,key-37} show that neutrinos
have a small but non-zero mass, suggesting the existence of both left-
and right-handed neutrinos. In order to formulate the quaternionic
Dirac equation for left- and right-handed neutrino, let us start with
the covariant form of the free Dirac equation as
\begin{align}
(i\hbar\gamma^{\mu}\partial_{\mu}-m_{0}c)\,\psi & \,\,=\,\,0\,,\label{eq:17}
\end{align}
where $m_{0}$ is the rest mass of the Dirac-like particle and $\psi$
is the spinor. Now, to generalize the free Dirac equation (\ref{eq:17})
in terms of quaternions, we express the relation between Dirac spin
matrices (i.e., $\alpha,\beta\,\text{and}\,\gamma-$matrices) to the
left and right-handed quaternionic elements \textcolor{black}{as given
in Table 2. }\textbf{}
\begin{table}[H]
\begin{doublespace}
\centering{}%
\begin{tabular}{ccc}
\hline 
Matrices & Connection with left-handed quaternions & Connection with right-handed quaternions\tabularnewline
\hline 
$\sigma_{j}$ & $\sigma_{j}\,\,:\longmapsto\,\,ie_{j}$ & $\sigma_{j}\,\,:\longmapsto\,\,-ie_{j}$\tabularnewline
$\alpha_{j}$ & $\alpha_{j}\,\,:\longmapsto\,\,\delta e_{j}$ & $\alpha_{j}\,\,:\longmapsto\,\,-\delta e_{j}$\tabularnewline
$\beta$ & $\beta\,\,:\longmapsto\,\,\kappa e_{0}$ & $\beta\,\,:\longmapsto\,\,\kappa e_{0}$\tabularnewline
$\gamma^{j}$ & $\gamma^{j}\,\,:\longmapsto\,\,\eta e_{j}$ & $\gamma^{j}\,\,:\longmapsto\,\,-\eta e_{j}$\tabularnewline
\hline 
\end{tabular}\caption{Mapping of Pauli and Dirac matrices with left and right-handed quaternions}
\end{doublespace}
\end{table}
Here, in Table 2, the quaternionic valued of Dirac matrices can be
represented in terms of $2\times2$ Pauli's matrices ($\sigma_{0},\,\sigma_{j}$
for $j=1,2,3$),
\begin{align*}
\alpha_{j} & :=\,\,\,\begin{pmatrix}0 & \sigma_{j}\\
\sigma_{j} & 0
\end{pmatrix}\,\,,\beta:=\,\,\,\begin{pmatrix}\sigma_{0} & 0\\
0 & -\sigma_{0}
\end{pmatrix}\,\,,\gamma^{j}:=\,\,\,\begin{pmatrix}0 & \sigma_{j}\\
-\sigma_{j} & 0
\end{pmatrix}\,\,,
\end{align*}
with matrices $\eta$ = $\left(\begin{array}{cc}
0 & i\\
-i & 0
\end{array}\right)$ ; $\delta$ = $\left(\begin{array}{cc}
0 & i\\
i & 0
\end{array}\right)$ ; $\kappa$ = $\left(\begin{array}{cc}
1 & 0\\
0 & -1
\end{array}\right)$. Obviously, the quaternionic representation of Dirac matrices retained
their usual identities essential to maintaining the relativistic behavior
of the Dirac equation, thus
\begin{align}
\alpha_{j}^{2}\,=\,\beta^{2}\, & =\,1\,,\,\,\,\,\alpha_{j}\beta+\beta\alpha_{j}=0\,,\nonumber \\
\alpha_{i}\alpha_{j}+\alpha_{j}\alpha_{i} & \,\,=0\,;\,\,\forall\,i\neq j\,;\,\,\,(i,j)=1,2,3\,.\label{eq:18}
\end{align}
Correspondingly, we can write the chirality operator $\gamma^{5}$
in terms of quaternionic scalar unit element as
\begin{equation}
\gamma^{5}\,\,=\,\,i\gamma^{0}\gamma^{1}\gamma^{2}\gamma^{3}\,\,=\,\,\rho e_{0}\,;\,\,\,\,\,\,\,\rho\,\,=\,\,\begin{pmatrix}0 & 1\\
1 & 0
\end{pmatrix}\,,\label{eq:19}
\end{equation}
where $\gamma^{0}=\,\,\begin{pmatrix}1 & 0\\
0 & -1
\end{pmatrix}$. Now, keeping in view the connection of Dirac matrices with left-handed
quaternions, we can establish the left-handed quaternionic Dirac equation
for neutrino as 
\begin{equation}
\left[e_{0}\left(\kappa p_{0}-m_{0}c\right)-\sum_{j=1}^{3}e_{j}\left(\eta p_{j}+m_{j}c\right)\right]\circ\Psi\,\,=\,\,0\,,\,\,\,\,\text{(Left handed Dirac equation)}\label{eq:20}
\end{equation}
where we substitute the Dirac term $i\hbar\gamma^{\mu}\partial_{\mu}=i\hbar(\gamma^{0}\partial_{0}+\gamma^{1}\partial_{1}+\gamma^{2}\partial_{3}+\gamma^{3}\partial_{3}),$
$=\gamma^{0}p_{0}-\sum_{j=1}^{3}\gamma^{j}p_{j}$ connected with four
momentum of the neutrino $\left\{ p_{0}=i\hbar\partial_{0},p_{j}=-i\hbar\partial_{j}\right\} $
in quaternionic four-space. Also, we have used quaternionic four-masses
of neutrino, i.e., $\mathbb{M}\left(m_{0},m_{j}\right)$ to incorporate
both rest and moving mass of neutrino. The quaternionic value of the
Dirac spinor ($\Psi$) can be expressed as
\begin{align}
\Psi & \,\,=\,\,e_{0}\psi_{0}+\sum_{j=1}^{3}e_{j}\psi_{j}\,=e_{0}(\psi_{0}+e_{1}\psi_{1})+e_{2}(\psi_{2}-e_{1}\psi_{3})\,\nonumber \\
\, & \,\,=\,\,\psi_{a}+e_{2}\psi_{b}\,,\label{eq:21}
\end{align}
where $\psi_{a}=\left(\psi_{0}+e_{1}\psi_{1}\right)$ and $\psi_{b}=\left(\psi_{2}-e_{1}\psi_{3}\right)$.
Thus, the quaternionic Dirac spinor (\ref{eq:21}) can also be expressed
in terms of four-component form as
\begin{equation}
\Psi^{L}:\longmapsto\Psi\,\,=\,\,\left(\begin{array}{c}
\psi_{a}\\
\psi_{b}
\end{array}\right)\,\,=\,\,\left(\begin{array}{c}
\psi_{0}+e_{1}\psi_{1}\\
\psi_{2}-e_{1}\psi_{3}
\end{array}\right)\,\,=\,\,\begin{pmatrix}\psi_{0}\\
\psi_{1}\\
\psi_{2}\\
-\psi_{3}
\end{pmatrix}\,.\label{eq:22}
\end{equation}
As result, we can obtain the right-handed quaternionic Dirac equation
for neutrino as
\begin{equation}
\left[e_{0}\left(\kappa p_{0}-m_{0}c\right)+\sum_{j=1}^{3}e_{j}\left(\eta p_{j}-m_{j}c\right)\right]\circ\Psi\,\,=\,\,0\,,\,\,\,\,\text{(Right handed Dirac equation)}\,\,\,.\label{eq:23}
\end{equation}
It is noted that the quaternionic valued spinor $\Psi$ will change
as the quaternionic handedness is changed. Thus, for a right-handed
spinor, the values of $\psi_{a}$ and $\psi_{b}$ become $\left(\psi_{0}+e_{1}\psi_{1}\right)$
and $\left(\psi_{2}+e_{1}\psi_{3}\right)$, respectively. Hence, the
right-handed quaternionic spinor $\Psi$ can then be expressed as
\begin{equation}
\Psi^{R}:\longmapsto\Psi\,\,=\,\,\left(\begin{array}{c}
\psi_{a}\\
\psi_{b}
\end{array}\right)\,\,=\,\,\left(\begin{array}{c}
\psi_{0}+e_{1}\psi_{1}\\
\psi_{2}+e_{1}\psi_{3}
\end{array}\right)\,\,=\,\,\begin{pmatrix}\psi_{0}\\
\psi_{1}\\
\psi_{2}\\
\psi_{3}
\end{pmatrix}\,.\label{eq:24}
\end{equation}
Using the left and right-handed spinors, we will discuss the solution
of the energy and momentum spinors of neutrino in terms of quaternionic
four-spaces.

\section{Quaternionic handedness for Dirac neutrinos}

Because the extended quaternionic Dirac equation consists of scalar
and vector coefficients, we can examine the spinor solution in terms
of both coefficients. The quaternionic scalar coefficient of $e_{0}$
contains the energy term, so the quaternionic spinor solutions will
represent the energy state of the neutrino. Furthermore, the quaternionic
vector coefficient of $e_{j}$ includes the momentum term, which represents
the neutrino's momentum state. However, the usual free Dirac equation
refused to allow the segregation of neutrino energy and momentum states
in a single frame.

\subsection{Left and right-handed quaternionic spinor solutions for energy representation}

To extract the energy solutions of left-handed neutrino from the left
quaternionic Dirac equation (\ref{eq:20}), as discussed before, we
take the coefficient of $e_{0}$ as
\begin{equation}
(\kappa p_{0}-m_{0}c)\circ\Psi\,=\,0\,,\label{eq:25}
\end{equation}
which gives
\begin{equation}
\left(\begin{array}{cc}
p_{0}-m_{0}c & 0\\
0 & -p_{0}-m_{0}c
\end{array}\right)\left(\begin{array}{c}
\psi_{a}\\
\psi_{b}
\end{array}\right)\,\,=\,\,\left(\begin{array}{c}
0\\
0
\end{array}\right)\,.\label{eq:26}
\end{equation}
Substituting the values of left-handed spinor components $\left\{ \psi_{a},\psi_{b}\right\} $
from equation (\ref{eq:21}), we obtain
\begin{align}
\psi_{2}\,=\,\left(\tfrac{p_{0}-m_{0}c}{p_{0}+m_{0}c}\right)\psi_{0}\,,\,\,\,\,\,\,\psi_{3} & \,=-\left(\tfrac{p_{0}-m_{0}c}{p_{0}+m_{0}c}\right)\psi_{1}\,.\label{eq:27}
\end{align}
So, the left-handed quaternionic Dirac spinor for neutrino can be
expressed in terms of energy components as
\begin{equation}
\Psi=\left(\begin{array}{c}
\psi_{0}+e_{1}\psi_{1}\\
\left(\tfrac{p_{0}-m_{0}c}{p_{0}+m_{0}c}\right)\psi_{0}-e_{1}\left[-\left(\tfrac{p_{0}-m_{0}c}{p_{0}+m_{0}c}\right)\psi_{1}\right]
\end{array}\right)=\left(\begin{array}{c}
\psi_{0}\\
\psi_{1}\\
\left(\tfrac{p_{0}-m_{0}c}{p_{0}+m_{0}c}\right)\psi_{0}\\
\left(\tfrac{p_{0}-m_{0}c}{p_{0}+m_{0}c}\right)\psi_{1}
\end{array}\right)\,.\label{eq:28}
\end{equation}
Correspondingly, using right-handed quaternionic Dirac equation, we
get the right-handed neutrino spinor $\Psi$ for energy representation
as
\begin{equation}
\Psi\,\,=\,\,\left(\begin{array}{c}
\psi_{0}+e_{1}\psi_{1}\\
\left(\tfrac{p_{0}-m_{0}c}{p_{0}+m_{0}c}\right)\psi_{0}+e_{1}\left[\left(\tfrac{p_{0}-m_{0}c}{p_{0}+m_{0}c}\right)\psi_{1}\right]
\end{array}\right)\,=\,\left(\begin{array}{c}
\psi_{0}\\
\psi_{1}\\
\left(\tfrac{p_{0}-m_{0}c}{p_{0}+m_{0}c}\right)\psi_{0}\\
\left(\tfrac{p_{0}-m_{0}c}{p_{0}+m_{0}c}\right)\psi_{1}
\end{array}\right)\,.\label{eq:29}
\end{equation}
Interestingly, the left and right-handed quaternionic spinors have
the same energy representation of a neutrino at rest mass ($m_{0})$.
This makes sense because in a quaternionic product, the handedness
does not change when multiplied by the scalar unit $e_{0}$. Now,
let us choose the standard spinor components as $\left(\begin{array}{c}
\psi_{0}\\
\psi_{1}
\end{array}\right)=$ $\left(\begin{array}{c}
1\\
0
\end{array}\right)$ or $\left(\begin{array}{c}
0\\
1
\end{array}\right)$, then the quaternionic energy spinor solutions for left or right-handed
neutrinos become:
\begin{equation}
\Psi^{\uparrow}\,\,=\,\,\left(\begin{array}{c}
1\\
0\\
\left(\tfrac{p_{0}-m_{0}c}{p_{0}+m_{0}c}\right)\\
0
\end{array}\right)\,;\,\,\,\,\,\,\,\,\Psi^{\downarrow}\,\,=\,\,\left(\begin{array}{c}
0\\
1\\
0\\
\left(\tfrac{p_{0}-m_{0}c}{p_{0}+m_{0}c}\right)
\end{array}\right)\,,\label{eq:30}
\end{equation}
where $\Psi^{\uparrow}$ and $\Psi^{\downarrow}$ are respectively
spin-up and spin-down spinors. Obviously, the quaternionic Dirac equation
also gives spinor solutions for the antineutrino. So, we can obtain
them by putting $p_{0}=-p_{0}$, which makes the negative energy like
spinor solutions for antineutrinos as stated by the Dirac sea interpretation.
Hence, they are given as
\begin{equation}
\Psi_{anti}^{\uparrow}\,\,=\,\,\left(\begin{array}{c}
\left(\tfrac{p_{0}-m_{0}c}{p_{0}+m_{0}c}\right)\\
0\\
1\\
0
\end{array}\right)\,;\,\,\,\,\,\,\,\,\,\,\Psi_{anti}^{\downarrow}\,\,=\,\,\left(\begin{array}{c}
0\\
\left(\tfrac{p_{0}-m_{0}c}{p_{0}+m_{0}c}\right)\\
0\\
1
\end{array}\right)\,.\label{eq:31}
\end{equation}
Furthermore, the quaternionic spinor solutions for neutrinos and antineutrinos
in equations (\ref{eq:30}) and (\ref{eq:31}) are found to be helicity
eigenstates, representing the energy states of neutrinos and antineutrinos'
mass eigenstates.

\subsection{Left and right-handed quaternionic spinor solutions for momentum
representation}

The momentum states \cite{key-18,key-22}, when considering the quaternionic
four-momentum in relativistic quantum physics, can allow a deeper
understanding of particle behavior and interactions. To obtain the
quaternionic momentum states for Dirac neutrinos, we take the coefficient
of $e_{j}$ in (\ref{eq:20}) as
\begin{equation}
\sum_{j=1}^{3}(\eta p_{j}+m_{j}c)\circ\Psi\,=\,0\,,\label{eq:32}
\end{equation}
which can be further written as
\begin{equation}
\left(\begin{array}{cc}
m_{j}c & ip_{j}\\
-ip_{j} & m_{j}c
\end{array}\right)\left(\begin{array}{c}
\psi_{a}\\
\psi_{b}
\end{array}\right)\,\,=\,\,\left(\begin{array}{c}
0\\
0
\end{array}\right)\,.\label{eq:33}
\end{equation}
Using the values of $\psi_{a}$ and $\psi_{b}$ in terms of left-handed
quaternionic realization, we get
\begin{align}
\psi_{2} & \,\,=-\left(\tfrac{m_{j}c-ip_{j}}{m_{j}c+ip_{j}}\right)\psi_{0}\,,\,\,\,\,\,\,\,\,\psi_{3}\,\,=\,\left(\tfrac{m_{j}c-ip_{j}}{m_{j}c+ip_{j}}\right)\psi_{1}\,,\label{eq:34}
\end{align}
which gives the quaternionic momentum spinor for left-handed neutrinos
as
\begin{equation}
\Psi^{L}\,\,=\,\,\left(\begin{array}{c}
\psi_{0}+e_{1}\psi_{1}\\
-\left(\tfrac{m_{j}c-ip_{j}}{m_{j}c+ip_{j}}\right)\psi_{0}-e_{1}\left(\tfrac{m_{j}c-ip_{j}}{m_{j}c+ip_{j}}\right)\psi_{1}
\end{array}\right)\,\,=\,\,\left(\begin{array}{c}
\psi_{0}\\
\psi_{1}\\
-\left(\tfrac{m_{j}c-ip_{j}}{m_{j}c+ip_{j}}\right)\psi_{0}\\
-\left(\tfrac{m_{j}c-ip_{j}}{m_{j}c+ip_{j}}\right)\psi_{1}
\end{array}\right)\,.\label{eq:35}
\end{equation}
In standard spinor notation, equation (\ref{eq:35}) can be reduced
as
\begin{equation}
\Psi_{+}^{L\uparrow}\,\,=\,\,\left(\begin{array}{c}
1\\
0\\
-\left(\tfrac{m_{j}c\,-\,ip_{j}}{m_{j}c\,+\,ip_{j}}\right)\\
0
\end{array}\right)\,;\,\,\,\,\,\,\Psi_{+}^{L\downarrow}\,\,=\,\,\left(\begin{array}{c}
0\\
1\\
0\\
-\left(\tfrac{m_{j}c-ip_{j}}{m_{j}c+ip_{j}}\right)
\end{array}\right)\,,\label{eq:36}
\end{equation}
where the superscript `$L$' denotes the left quaternionic handedness
of the spinor solutions. Further, we can also write the momentum spinor
representation for left-handed antineutrinos as
\begin{equation}
\Psi_{-}^{L\uparrow}\,\,=\,\,\left(\begin{array}{c}
-\left(\tfrac{m_{j}c-ip_{j}}{m_{j}c+ip_{j}}\right)\\
0\\
1\\
0
\end{array}\right)\,;\,\,\,\,\,\,\,\,\Psi_{-}^{L\downarrow}\,\,=\,\,\left(\begin{array}{c}
0\\
-\left(\tfrac{m_{j}c-ip_{j}}{m_{j}c+ip_{j}}\right)\\
0\\
1
\end{array}\right)\,.\label{eq:37}
\end{equation}
Here, the antineutrinos can assume to propagate backwards in time,
making their energy and momentum negative. For right-handed particles,
we can use the right-handed quaternionic multiplication. Thus the
momentum spinor solutions for the right-handed quaternionic Dirac
equation are obtained as
\begin{equation}
\Psi^{R}\,\,=\,\,\left(\begin{array}{c}
\psi_{0}+e_{1}\psi_{1}\\
-\left(\tfrac{m_{j}c+ip_{j}}{m_{j}c-ip_{j}}\right)\psi_{0}+e_{1}\left[-\left(\tfrac{m_{j}c+ip_{j}}{m_{j}c-ip_{j}}\right)\psi_{1}\right]
\end{array}\right)\,\,=\,\,\left(\begin{array}{c}
\psi_{0}\\
\psi_{1}\\
-\left(\tfrac{m_{j}c+ip_{j}}{m_{j}c-ip_{j}}\right)\psi_{0}\\
-\left(\tfrac{m_{j}c+ip_{j}}{m_{j}c-ip_{j}}\right)\psi_{1}
\end{array}\right)\,.\label{eq:38}
\end{equation}
Therefore, the quaternionic momentum spinor solutions for right-handed
neutrinos and antineutrinos can be expressed by
\begin{equation}
\Psi_{+}^{R\uparrow}\,\,=\,\,\left(\begin{array}{c}
1\\
0\\
-\left(\tfrac{m_{j}c+ip_{j}}{m_{j}c-ip_{j}}\right)\\
0
\end{array}\right)\,;\,\,\,\,\,\,\Psi_{+}^{R\downarrow}\,\,=\,\,\left(\begin{array}{c}
0\\
1\\
0\\
-\left(\tfrac{m_{j}c+ip_{j}}{m_{j}c-ip_{j}}\right)
\end{array}\right)\,,\label{eq:39}
\end{equation}
\begin{equation}
\Psi_{-}^{R\uparrow}\,\,=\,\,\left(\begin{array}{c}
-\left(\tfrac{m_{j}c+ip_{j}}{m_{j}c-ip_{j}}\right)\\
0\\
1\\
0
\end{array}\right)\,;\,\,\,\,\,\,\,\Psi_{-}^{R\downarrow}\,\,=\,\left(\begin{array}{c}
0\\
-\left(\tfrac{m_{j}c+ip_{j}}{m_{j}c-ip_{j}}\right)\\
0\\
1
\end{array}\right)\,.\label{eq:40}
\end{equation}
It is noted that all four quaternionic momentum spinor solutions in
both handedness of quaternion are also eigenstates of helicity. Interestingly,
the complex conjugate of a quaternionic momentum spinor state is equivalent
to changing its handedness, \emph{i.e.,}$\Psi^{R}\,=\,(\Psi^{L})^{*}\,,\,\Psi^{L}=(\Psi^{R})^{*}\,.$

\section{Quaternionic handedness for Majorana neutrinos}

Since the concept of Majorana neutrinos \cite{key-5,key-6} extends
beyond the traditional framework of Dirac neutrinos. The distinction
between Majorana and Dirac neutrinos lies in whether neutrinos are
their own antiparticles \cite{key-38}. The Majorana neutrinos, therefore,
fulfill the condition $\psi=\psi^{C};\,\psi^{C}=i\gamma^{2}\psi^{*}$,
where $C$ is the charge conjugation operator for a particle that
converts a particle state into an anti-particle.

\subsection{Quaternionic energy spinor solution}

The spinor solutions of the Majorana neutrinos can be found by linearly
superposing all spinor solutions of the Dirac equation\textbf{ }and
equating them with their own charge conjugate \cite{key-39}. Thus,
the generalized quaternionic Majorana spinor ($\Phi_{M}$) for energy
state representation can be obtained as
\begin{align}
\Phi_{M} & :\longmapsto\Phi\,=\,a_{1}\Psi_{+}^{\uparrow}+a_{2}\Psi_{+}^{\downarrow}+a_{3}\Psi_{-}^{\downarrow}+a_{4}\Psi_{-}^{\uparrow}\,,\label{eq:41}
\end{align}
such that,
\begin{align}
\Phi\,\longmapsto\,\,\left(\begin{array}{c}
\phi_{0}\\
\phi_{1}\\
\phi_{2}\\
\phi_{3}
\end{array}\right) & \,=\,\,\begin{pmatrix}a_{1}+a_{4}\left(\tfrac{p_{0}-m_{0}c}{p_{0}+m_{0}c}\right)\\
a_{2}+a_{3}\left(\tfrac{p_{0}-m_{0}c}{p_{0}+m_{0}c}\right)\\
a_{1}\left(\tfrac{p_{0}-m_{0}c}{p_{0}+m_{0}c}\right)+a_{4}\\
a_{2}\left(\tfrac{p_{0}-m_{0}c}{p_{0}+m_{0}c}\right)+a_{3}
\end{pmatrix}\,,\label{eq:42}
\end{align}
where $a_{1},\,a_{2},\,a_{3}$ and $a_{4}$ are arbitrary coefficients
of Dirac states. Conditionally, a quaternionic Majorana spinor will
be equal to its own charge conjugate if,
\begin{align}
\phi_{3} & \,=\,\phi_{0}^{*}\,,\,\,\,\,\,\,\,\phi_{2}\,=-\phi_{1}^{*}\,.\label{eq:43}
\end{align}
Henceforth, by using the quaternionic Majorana condition (\ref{eq:43}),
we may obtain 
\begin{alignat}{1}
\frac{a_{1}^{*}-a_{3}}{a_{2}-a_{4}^{*}} & \,\,=\,\,\left(\tfrac{p_{0}-m_{0}c}{p_{0}+m_{0}c}\right)\,,\,\,\,\,\,\,\,\,\,\frac{-a_{2}^{*}-a_{4}}{a_{1}+a_{3}^{*}}\,\,=\,\,\left(\tfrac{p_{0}-m_{0}c}{p_{0}+m_{0}c}\right)\,\,,\label{eq:44}
\end{alignat}
which may give $a_{1}^{*}=p_{0}$ ; $a_{2}^{*}=m_{0}c$ ; $a_{3}=m_{0}c$
and $a_{4}=-p_{0}$. Thus, from equation (\ref{eq:42}), the possible
quaternionic energy spinor representation for Majorana neutrinos can
be expressed as
\begin{equation}
\Phi\,\longmapsto\,\Phi_{p_{0}}^{1}\,\,=\,\,N\left(\begin{array}{c}
1\\
1\\
-1\\
1
\end{array}\right)\,,\label{eq:45}
\end{equation}
where $N=\tfrac{2p_{0}m_{0}c}{p_{0}+m_{0}c}$. Further, from equation
(\ref{eq:44}), the other possible energy spinor solution can be obtained
by equating the coefficient ratio as \emph{$\frac{a_{1}^{*}-a_{3}}{a_{2}-a_{4}^{*}}=\frac{-a_{2}^{*}-a_{4}}{a_{1}+a_{3}^{*}}\,.$}
Assuming arbitrary coefficients to be real for simplicity, we found
$a_{2}=-\left(\tfrac{p_{0}}{m_{0}c}\right)a_{1}$ and $a_{3}=\left(\tfrac{p_{0}}{m_{0}c}\right)a_{1}$,
which leads to the another possible Majorana energy spinor solution
as
\begin{equation}
\Phi\,\longmapsto\,\Phi_{p_{0}}^{2}\,\,=\,\,N\left(\begin{array}{c}
1\\
-1\\
1\\
1
\end{array}\right)\,.\label{eq:46}
\end{equation}
Hence, we obtained a couple of quaternionic Majorana states ($\Phi_{p_{0}}^{1},\Phi_{p_{0}}^{2}$)
for energy representation of neutrinos that are orthogonal and invariant
under charge conjugation. It should also be noted that handedness
is not taken into account for the quaternionic scalar quantities associated
with the spinor representation.

\subsection{Quaternionic left and right-handed momentum spinor solution}

In order to obtain left-handed quaternionic momentum spinor solutions
for Majorana neutrinos, we begin with the generalized spinor superimposed
by the quaternionic Dirac momentum spinor solutions, as
\begin{align}
\Phi_{p}^{L} & \,\,=\,\,a_{1}\Psi_{p+}^{L\uparrow}+a_{2}\Psi_{p+}^{L\downarrow}+a_{3}\Psi_{p-}^{L\downarrow}+a_{4}\Psi_{p-}^{L\uparrow}\,,\label{eq:47}
\end{align}
which yields
\begin{align}
\Phi_{p}^{L}:\longmapsto\,\left(\begin{array}{c}
\phi_{0}\\
\phi_{1}\\
\phi_{2}\\
\phi_{3}
\end{array}\right)\,\,=\,\,\begin{pmatrix}a_{1}-a_{4}\left(\tfrac{m_{j}c\,-\,ip_{j}}{m_{j}c\,+\,ip_{j}}\right)\\
a_{2}-a_{3}\left(\tfrac{m_{j}c\,-\,ip_{j}}{m_{j}c\,+\,ip_{j}}\right)\\
-a_{1}\left(\tfrac{m_{j}c\,-\,ip_{j}}{m_{j}c\,+\,ip_{j}}\right)+a_{4}\\
-a_{2}\left(\tfrac{m_{j}c\,-\,ip_{j}}{m_{j}c\,+\,ip_{j}}\right)+a_{3}
\end{pmatrix}\, & .\label{eq:48}
\end{align}
Applying the condition (\ref{eq:43}) in equation (\ref{eq:48}),
then we get the following relations,
\begin{align}
a_{4}^{*}\,\,=-a_{2}\,\,\,\mathrm{and}\,\,\,\frac{a_{1}^{*}-a_{3}}{a_{4}^{*}-a_{2}} & \,\,=\,\,\tfrac{m_{j}^{2}c^{2}\,-\,p_{j}^{2}}{m_{j}^{2}c^{2}\,+\,p_{j}^{2}}\,,\label{eq:49}\\
a_{1}\,\,=a_{3}^{*}\,\,\,\mathrm{and\,\,\,}\frac{a_{4}^{*}+a_{2}}{a_{3}^{*}+a_{1}} & \,\,=\,\,\tfrac{m_{j}^{2}c^{2}\,-\,p_{j}^{2}}{m_{j}^{2}c^{2}\,+\,p_{j}^{2}}\,.\label{eq:50}
\end{align}
Solving equations (\ref{eq:49}) and (\ref{eq:50}), we obtain the
condition $p_{j}=\pm m_{j}c\,,$which gives $\tfrac{m_{j}c\,-\,ip_{j}}{m_{j}c\,+\,ip_{j}}=\tfrac{1\,\mp\,i}{1\,\pm\,i}$.
Thus, the left-handed quaternionic momentum spinor for Majorana neutrinos
is expressed by
\begin{equation}
\Phi_{p}^{L}\,\,=\,\,\begin{pmatrix}a_{1}+a_{2}^{*}\left(\tfrac{1\,\mp\,i}{1\,\pm\,i}\right)\\
a_{2}-a_{1}^{*}\left(\tfrac{1\,\mp\,i}{1\,\pm\,i}\right)\\
-a_{1}\left(\tfrac{1\,\mp\,i}{1\,\pm\,i}\right)-a_{2}^{*}\\
-a_{2}\left(\tfrac{1\,\mp\,i}{1\,\pm\,i}\right)+a_{1}^{*}
\end{pmatrix}\,.\label{eq:51}
\end{equation}
As such, the right-handed quaternionic momentum spinor for right-handed
neutrinos become
\begin{equation}
\Phi_{p}^{R}\,\,=\,\,\begin{pmatrix}a_{1}+a_{2}^{*}\left(\tfrac{1\,\pm\,i}{1\,\mp\,i}\right)\\
a_{2}-a_{1}^{*}\left(\tfrac{1\,\pm\,i}{1\,\mp\,i}\right)\\
-a_{1}\left(\tfrac{1\,\pm\,i}{1\,\mp\,i}\right)-a_{2}^{*}\\
-a_{2}\left(\tfrac{1\,\pm\,i}{1\,\mp\,i}\right)+a_{1}^{*}
\end{pmatrix}\,.\label{eq:52}
\end{equation}
So, it is analyzed that for any arbitrary value of $a_{1}$ and $a_{2}$,
the Majorana momentum spinors $\Phi_{p}^{L}$ and $\Phi^{R}$ are
invariant under charge conjugation using the identity $\tfrac{1\,\mp\,i}{1\,\pm\,i}=-\left(\tfrac{1\,\pm\,i}{1\,\mp\,i}\right)$.
Also, the charge conjugation of quaternionic Majorana momentum spinors
can be defined in terms of their chiral spinors, such as $(\Phi_{p}^{L})^{C}=i\eta e_{2}\Phi_{p}^{R}\,,\text{\,and }(\Phi_{p}^{R})^{C}=-i\eta e_{2}\Phi_{p}^{L}\,.$

\section{Quaternionic flavor oscillations of massive Dirac neutrinos}

The mass eigenstates of Dirac and Majorana neutrinos are the eigenstates
of Hamiltonian having definite mass. The mass eigenstates in terms
of quaternionic algebra will satisfy quaternionic Dirac equations
(\ref{eq:20}) and (\ref{eq:23}). Let us consider the two flavors
of neutrino oscillations defined by the rotational transformation
as \cite{key-40},
\begin{equation}
\begin{pmatrix}\nu_{e}(0)\\
\nu_{\mu}(0)
\end{pmatrix}\,\,=\,\,\begin{pmatrix}\cos\theta & \sin\theta\\
-\sin\theta & \cos\theta
\end{pmatrix}\begin{pmatrix}\nu_{1}(0)\\
\nu_{2}(0)
\end{pmatrix}\,,\label{eq:53}
\end{equation}
where $\nu_{e}(0),\nu_{\mu}(0)$ are electron and muon flavour states
at time $t=0$, respectively; $\theta$ is the mixing angle for mass
eigenstates $\nu_{1}(0)$ and $\nu_{2}(0)$. Since the mass-eigenstates
are necessarily orthonormal \cite{key-41}, we can choose the quaternionic
spinor solutions as the mass eigenstates with different spin orientations.
Thus, the mass-eigenstates can be written in terms of quaternionic
energy spinors as
\begin{equation}
\nu_{1}(0):\longmapsto\Psi_{1}^{\uparrow}\,=\,K_{1}\left(\begin{array}{c}
1\\
0\\
\left(\tfrac{p_{0}-cm_{0}^{\nu_{1}}}{p_{0}+cm_{0}^{\nu_{1}}}\right)\\
0
\end{array}\right)\,;\,\,\,\nu_{2}(0):\longmapsto\Psi_{2}^{\downarrow}\,=\,K_{2}\left(\begin{array}{c}
0\\
1\\
0\\
\left(\tfrac{p_{0}-cm_{0}^{\nu_{2}}}{p_{0}+cm_{0}^{\nu_{2}}}\right)
\end{array}\right)\,,\label{eq:54}
\end{equation}
where $\Psi_{1}^{\uparrow}$ and $\Psi_{2}^{\downarrow}$ are the
spin-up and spin-down quaternionic energy spinors representing Dirac
neutrinos of rest masses $m_{0}^{\nu_{1}}$ and $m_{0}^{\nu_{2}}$,
respectively, with normalization constants $K_{1}=\left[1+\left(\tfrac{p_{0}-cm_{0}^{\nu_{1}}}{p_{0}+cm_{0}^{\nu_{1}}}\right)^{2}\right]^{\nicefrac{-1}{2}}$
and $K_{2}=\left[1+\left(\tfrac{p_{0}-cm_{0}^{\nu_{2}}}{p_{0}+cm_{0}^{\nu_{2}}}\right)^{2}\right]^{\nicefrac{-1}{2}}$.
The stationary state of quaternionic electron flavor-eigenstate of
neutrino $\left|\nu_{e}(0)\right\rangle $, and the time evolved quaternionic
muon flavor-eigenstate of neutrino $\left|\nu_{\mu}(t)\right\rangle $
are expressed by 
\begin{align}
\left|\nu_{e}(0)\right\rangle  & \,\,=\,\,\left|\Psi_{1}^{\uparrow}\right\rangle \cos\theta+\left|\Psi_{2}^{\downarrow}\right\rangle \sin\theta\,,\label{eq:55}\\
\mathrm{and,}\,\,\,\left|\nu_{\mu}(t)\right\rangle  & \,\,=\,-\left|\Psi_{1}^{\uparrow}\right\rangle \exp\left(-i\phi_{1}\right)\sin\theta+\left|\Psi_{2}^{\downarrow}\right\rangle \exp\left(-i\phi_{2}\right)\cos\theta\,.\label{eq:56}
\end{align}
Here $\phi_{l}=\left(\overrightarrow{p}\cdot\overrightarrow{x}-\frac{i}{\hslash}E^{\nu_{l}}t\right)\,;\,l=\left(1,\,2\right)$,
and $E^{\nu_{l}}=\sqrt{p^{2}c^{2}+(m_{0}^{\nu_{l}})^{2}c^{4}}$ is
the relativistic energy of the respective mass-eigenstate of neutrinos.
Moreover, the state $\left|\nu_{e}(0)\right\rangle $ describes the
initial energy state of the electron neutrino, which has undergone
weak interaction, while $\left|\nu_{\mu}(t)\right\rangle $ is the
energy state of the muon neutrino after experiencing neutrino oscillation.
Since, $\left|\Psi_{1}^{\uparrow}\right\rangle $ and $\left|\Psi_{2}^{\downarrow}\right\rangle $
are orthogonal, we can obtain the conventional transition probability
for the neutrino oscillation $(\nu_{e}\rightarrow\nu_{\mu})$ as $\mathcal{P}(\nu_{e}\rightarrow\nu_{\mu})=\left|\left\langle \nu_{e}(0)|\nu_{\mu}(t)\right\rangle \right|^{2}$
$=\sin^{2}\theta\cos^{2}\theta\left[2-\exp i\left(\phi_{2}-\phi_{1}\right)-\exp i\left(\phi_{1}-\phi_{2}\right)\right]\,.$
On the other hand, the normalized mass eigenstates of left or right-handed
neutrinos can be expressed in terms of their momentum-valued Dirac
spinors as
\begin{equation}
\nu_{1}(0):\longmapsto\Psi_{1}^{\left(L/R\right)\uparrow}\,\,=\,\,\frac{1}{\sqrt{2}}\left(\begin{array}{c}
1\\
0\\
-\left(\tfrac{cm_{j}^{\nu_{1}}\mp ip_{j}}{cm_{j}^{\nu_{1}}\pm ip_{j}}\right)\\
0
\end{array}\right)\,;\,\,\,\,\nu_{2}(0):\longmapsto\Psi_{2}^{\left(L/R\right)\downarrow}\,\,=\,\,\frac{1}{\sqrt{2}}\left(\begin{array}{c}
0\\
1\\
0\\
-\left(\tfrac{cm_{j}^{\nu_{2}}\mp ip_{j}}{cm_{j}^{\nu_{2}}\pm ip_{j}}\right)
\end{array}\right)\,,\label{eq:57}
\end{equation}
where $\Psi_{1}^{\left(L/R\right)\uparrow}$ and $\Psi_{2}^{\left(L/R\right)\downarrow}$
are the quaternionic left or right momentum spinors having definite
moving quaternionic masses $m_{j}^{\nu_{1}}$ and $m_{j}^{\nu_{2}}$,
respectively. The quaternionic formalism has the advantage of allowing
us to consider new mass-eigenstates for Dirac neutrinos arising from
the presence of momentum spinors. It should be noted that due to their
orthogonality in quaternionic mass-eigenstates, the transition probability
becomes the same.

\section{Quaternionic flavor oscillations of massive Majorana neutrinos}

In this part, we will extend the quaternionic-based Dirac formalism
of massive neutrinos flavor to the massive Majorana neutrinos flavor.
Thus, for Majorana neutrinos, the mixing matrix changes with an additional
Majorana phase $\varphi\longmapsto\begin{pmatrix}1 & 0\\
0 & e^{i\varphi}
\end{pmatrix}$ as \cite{key-12},
\begin{align}
\begin{pmatrix}\nu_{e}(0)\\
\nu_{\mu}(0)
\end{pmatrix} & \,\,=\,\,\begin{pmatrix}\cos\theta & \sin\theta\\
-\sin\theta & \cos\theta
\end{pmatrix}\begin{pmatrix}1 & 0\\
0 & e^{i\varphi}
\end{pmatrix}\begin{pmatrix}\nu_{1}(0)\\
\nu_{2}(0)
\end{pmatrix}\,,\nonumber \\
 & =\,\,\begin{pmatrix}\cos\theta & e^{i\varphi}\sin\theta\\
-\sin\theta & e^{i\varphi}\cos\theta
\end{pmatrix}\begin{pmatrix}\nu_{1}(0)\\
\nu_{2}(0)
\end{pmatrix}\,.\label{eq:58}
\end{align}
Now, we can found the normalized Majorana energy spinors $\Phi_{1}$
and $\Phi_{2}$ as
\begin{equation}
\nu_{1}(0):\longmapsto\Phi_{1}\,\,=\,\,\frac{1}{2}\begin{pmatrix}1\\
1\\
-1\\
1
\end{pmatrix}\,;\,\,\,\,\,\,\,\,\,\nu_{2}(0):\longmapsto\Phi_{2}\,\,=\,\,\frac{1}{2}\begin{pmatrix}1\\
-1\\
1\\
1
\end{pmatrix}\,.\label{eq:59}
\end{equation}
From equation (\ref{eq:58}), we can express the neutrinos states
$\nu_{e}(0)$ and $\nu_{\mu}(t)$ as
\begin{align}
\left|\nu_{e}(0)\right\rangle  & \,\,=\,\,\left|\Phi_{1}\right\rangle \cos\theta+\left|\Phi_{2}\right\rangle e^{i\varphi}\sin\theta\,,\label{eq:60}\\
\left|\nu_{\mu}(t)\right\rangle  & \,\,=\,-\left|\Phi_{1}\right\rangle \exp\left(-i\phi_{1}\right)\sin\theta+e^{i\varphi}\left|\Phi_{2}\right\rangle \exp\left(-i\phi_{2}\right)\cos\theta\,.\label{eq:61}
\end{align}
As such, we can construct left-handed two orthogonal momentum quaternionic
spinor solutions for Majorana neutrinos from equation (\ref{eq:51})
by putting $a_{1}=1$ ; $a_{2}=0$ and $a_{1}=0$ ; $a_{2}=1$ as
\begin{equation}
\nu_{1}(0):\longmapsto\Phi_{1}^{L}\,\,=\,\,\frac{1}{2}\begin{pmatrix}1\\
-\left(\tfrac{1\,\mp\,i}{1\,\pm\,i}\right)\\
-\left(\tfrac{1\,\mp\,i}{1\,\pm\,i}\right)\\
1
\end{pmatrix}\,;\,\,\,\,\,\,\,\,\nu_{2}(0):\longmapsto\Phi_{2}^{L}\,\,=\,\,\frac{1}{2}\begin{pmatrix}\left(\tfrac{1\,\mp\,i}{1\,\pm\,i}\right)\\
1\\
-1\\
-\left(\tfrac{1\,\mp\,i}{1\,\pm\,i}\right)
\end{pmatrix}\,,\label{eq:62}
\end{equation}
and for the right-handed quaternionic formalism, we can yield
\begin{equation}
\nu_{1}(0):\longmapsto\Phi_{1}^{R}\,\,=\,\,\frac{1}{2}\begin{pmatrix}1\\
-\left(\tfrac{1\,\pm\,i}{1\,\mp\,i}\right)\\
-\left(\tfrac{1\,\pm\,i}{1\,\mp\,i}\right)\\
1
\end{pmatrix}\,;\,\,\,\,\,\,\,\,\,\,\nu_{2}(0):\longmapsto\Phi_{2}^{R}\,\,=\,\,\frac{1}{2}\begin{pmatrix}\left(\tfrac{1\,\pm\,i}{1\,\mp\,i}\right)\\
1\\
-1\\
-\left(\tfrac{1\,\pm\,i}{1\,\mp\,i}\right)
\end{pmatrix}\,.\label{eq:63}
\end{equation}
It needs to be noted that the probability obtained from flavor eigenstates
established from momentum quaternionic spinors for both Dirac and
Majorana neutrinos is independent of the moving masses of the Majorana
neutrinos. This is critical for proving the substantial characteristics
of both Dirac and Majorana neutrinos since, in the case of such a
dependence, even moving massless particles with moving masses would
have shown neutrino oscillations. If we set the rest masses $m_{0}^{\nu_{1}}$
and $m_{0}^{\nu_{2}}$ equal to zero, then $\phi_{2}$ becomes equal
to $\phi_{1}$ because $E^{\nu_{1}}$ and $E^{\nu_{2}}$ are equal.
Thus, the transition probabilities obtained by adopting quaternionic
energy and momentum spinors as mass eigenstates for both Dirac and
Majorana neutrinos in both quaternionic chiralities, become $P(\nu_{e}\rightarrow\nu_{\mu})=\sin^{2}\theta\cos^{2}\theta\left[2-\exp i\left(0\right)-\exp i\left(0\right)\right]=0\,.$
This suggests that the probability of neutrino oscillations is non-zero
only if the neutrinos' rest masses and mass eigenstates are non-zero,
i.e., neutrino oscillations occur only when the neutrino is massive.
This implies that the right chiral neutrinos should exist, even if
they do not participate in weak interactions.

\section{Conclusion}

The research work substantiated the massive nature of the neutrinos
through the two flavor neutrino oscillations. The Dirac equation has
been formulated in both-handedness of the quaternionic algebra along
with the incorporation of four quaternionic masses, which generate
quaternionic spinor solutions depicting energy and momentum states
of left and right-handed Dirac neutrinos. The role of quaternionic
basis elements ($e_{0},e_{j}$) has been shown to be very important
to discussing the unified structure of quaternionic energy and momentum
spinor solutions. We have constructed the formulation of quaternionic
Majorana spinors that depict energy and momentum states for left and
right-handed Majorana neutrinos. Since they are not found to be eigenstates
of helicity but are also viable choices for mass eigenstates for Majorana
neutrinos. The transition probabilities have been calculated from
both quaternionic energy and momentum mass eigenstates, which match
the conventional oscillation probability, and concluded that the rest
masses of the neutrinos should be non-zero for occurring the oscillations.
However, we can further extend the two flavors to the three flavor
oscillation formulation, maintaining the quaternionic structure. Nevertheless,
the two flavor oscillations, which are actually exhibited by solar
and atmospheric neutrinos and hence are physically relevant. Furthermore,
we have found that the quaternionic transition probability of neutrino
oscillations is not zero if and only if the mass-eigenstates and rest
masses of the neutrinos are not zero. Therefore, with the help of
quaternionic left and right-handed formalism, we concluded that the
left-handed neutrinos, with their spin oriented in the opposite direction
to their momentum, are used in weak nuclear reactions, whereas the
right-handed neutrinos, with their weakly interacting nature and potential
stability, could be candidates for dark matter, cosmic rays, matter-antimatter
symmetry, and so on.

\end{document}